\documentclass{article}

\usepackage{arxiv}
\usepackage{amssymb}
\usepackage{graphicx}
\usepackage{amsthm}
\usepackage{amsmath}
\usepackage{verbatim}      
\usepackage[colorlinks]{hyperref}

\newcommand{\squeezeupt}{\vspace{-3mm}}

\title{\texttt{ElasticMatrix}: A MATLAB Toolbox for Anisotropic Elastic Wave Propagation in Layered Media}

\author{
  Danny R. Ramasawmy \\
  Department of Medical Physics\\ 
  and Biomedical Engineering\\
  University College London\\
  London, UK \\
   \And
  Ben T. Cox \\
  Department of Medical Physics \\
  and Biomedical Engineering\\
  University College London\\
  London, UK \\
  \And
  Bradley E. Treeby\\
  Department of Medical Physics \\
  and Biomedical Engineering\\
  University College London\\
  London, UK \\
}

\begin{document}
\maketitle

\begin{abstract}
Simulating the propagation of elastic waves in multi-layered media has many applications. A common approach is to use matrix methods where the elastic wave-field within each material layer is represented by a sum of partial-waves along with boundary conditions imposed at each interface. While these methods are well-known, coding the required matrix formation, inversion, and analysis for general multi-layered systems is non-trivial and time-consuming. Here, a new open-source toolbox called \texttt{ElasticMatrix} is described which solves the problem of acoustic and elastic wave propagation in multi-layered media for isotropic and transverse-isotropic materials where the wave propagation occurs in a material plane of symmetry. The toolbox is implemented in MATLAB using an object oriented programming framework and is designed to be easy to use and extend. Methods are provided for calculating and plotting dispersion curves, displacement and stress fields, reflection and transmission coefficients, and slowness profiles.
\end{abstract}




\section{Motivation and Significance}
\label{section:motivation}

    Matrix models of wave propagation in multi-layered elastic solids have had a significant contribution to research areas such as acoustics, geophysics and electromagnetics.  
    A few examples include: structural health monitoring \cite{willberg2015simulation}, characterisation of interface bonding \cite{song2017characterization}, detection of debonding in joints \cite{rucka2019detection}, measuring material properties \cite{Lowe1995a}, designing composite layered structures \cite{nayfeh1995wave},  mode sorting of guided waves \cite{hakoda2019using}, the physical interpretation of guided wave structures \cite{hakoda2018using}, modelling the directional response of Fabry-P\'erot ultrasound sensors \cite{ramasawmy2018directivity}, reflection and transmission of plane waves \cite{chen2017reflection}, elastography of layered soft tissues \cite{li2019guided}, and ice detection on wind turbines \cite{shoja2015investigating}. 
    
    Matrix methods, in particular the partial-wave and global matrix method, represent the stress and displacement fields as a sum of partial-waves for each material of the layered-structure. Each partial-wave represents an upward or downward travelling (quasi-)compressional or (quasi-)shear wave. By invoking boundary conditions at the interfaces of adjacent layers, the partial-wave amplitudes and field properties of the first layer can be related to the last in the form of a `global' matrix. The resulting matrix equation can be used in two different ways. Firstly, the roots of the equation can be found which give the modal solutions or dispersion curves. Secondly, a subset of partial-wave amplitudes can be defined and the remaining amplitudes solved for. This can be used to calculate the displacement and stress fields within the multi-layered structure when a plane wave is incident. This method will be discussed further in Section \ref{section:modelDescription}. 
    
    Despite its usefulness, there are few available implementations of the partial-wave method. The current state-of-the-art implementation is Disperse \cite{pavlakovic1997disperse}. This software has been in development since 1990 and is primarily focused on calculating the dispersion solutions for multi-layered structures. 
    The Disperse software was originally based the partial-wave method, however, is currently being updated to use the spectral collocation method \cite{pavlakovic1997disperse,quintanilla2016full, quintanilla2015guided, quintanilla2015modeling, adamou2004spectral}. The main limitation with Disperse is that it is closed-source. For this reason it is not easily adaptable for applications that are not dispersion analysis, for example, extracting reflection coefficients or slowness profiles. There currently is only a single open-source code modelling the partial-wave method, (LAMB \cite{prego2010lamb}), however, this is limited to modelling only an isotropic plate.
    
    In this paper, a new open-source toolbox called \texttt{ElasticMatrix} is introduced which uses the partial-wave method for multi-layered structures with an arbitrary number of isotropic and transverse-isotropic materials. Where possible, it is validated against existing literature and has been implemented so that it is both easy to use and extend. Some potential uses of this software are: 1) plotting the slowness profiles of materials, 2) determining the reflection and transmission coefficients of multi-layered structures, 3) finding the dispersion curves of multi-layered structures, 4) plotting the displacement and stress fields, 5) extending the toolbox for other applications, for example modelling the directional response of Fabry-Perot ultrasound sensors \cite{ramasawmy2018directivity, cox2007frequency}. 
	
	A brief overview of the underlying mathematical model is described in Section \ref{section:modelDescription}. A selection of code snippets and examples are shown in Section \ref{section:softwareDescription}. (More extensive examples are available with the toolbox documentation.) The impact and conclusions are described in Section \ref{section:impact}.

  	\begin{figure}
        \centering
        \includegraphics[width=0.6\textwidth]{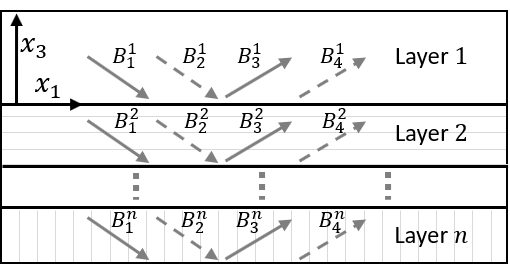}
        \caption{Diagram for an $n$-layered elastic medium. In the 2D plane there are four partial-waves with amplitude $B_i^n$, these represent (quasi-)compressional (solid arrows) and (quasi-)shear (dashed arrows) waves travelling upwards and downwards in each layer.}
        \label{fig:workingDiagram}
    \end{figure}

\section{Model Description}
\label{section:modelDescription}
    \subsection{Overview}
    \label{section:ElasticOverview}
    \texttt{ElasticMatrix} uses the partial-wave method to model wave propagation in multi-layered elastic solids. The method describes elastic plane-wave propagation along a plane of symmetry for n-layers of rigidly bonded transverse-isotropic materials. 
    An example of an isotropic material is glass, where the material properties are the same when measured from every direction. An example of a transverse-isotropic material is a bundle of fibres, where the properties have translational symmetry axially along the fibre, and are isotropic in the plane perpendicular to this. 
    \texttt{ElasticMatrix} can model layered transverse-isotropic materials if they are aligned such that they have rotational symmetry about the axis perpendicular to the plane of each layer and the wave-vector of the propagating wave lies in the plane of symmetry. In this case, the multi-layered structure can be modelled in two-dimensions. This is illustrated in Fig. \ref{fig:workingDiagram}. 

    Each partial-wave represents the superposition of waves that have been multiply reflected or transmitted at the interfaces between each layer in a steady-state. The polarisation vector and wave-vector of each of these partial-waves can be found from the Christoffel Equation which is described in Section \ref{section:unbound}. The degree of reflection and transmission depends on the boundary conditions at the interfaces and material properties of each layer. The coupled equations that arise from the boundary conditions can be combined into a `global-matrix' which allows them to be solved simultaneously, which is discussed in Section \ref{section:layered}. This global matrix approach can be used to tackle various problems in elastic wave propagation. For example, the singularities of the global matrix give the dispersion curves, and by specifying an incident wave, the resulting wave-field throughout the structure can be calculated. More detailed descriptions of the partial-wave and global-matrix method can be found in \cite{Lowe1995a, nayfeh1995wave, thomson1950transmission, nayfeh1991general, solie1973elastic, Rose2004,  haskell1953dispersion, brekhovskikh2012waves}.
    
    \subsection{Wave-vectors and Polarisation}
    \label{section:unbound}
    Firstly, the solution for a plane wave propagating in an unbounded medium is derived. This is needed to calculate the polarisation and wave-vectors for each partial-wave component and the process is repeated independently for every layer.
	The wave-equation for an anisotropic unbounded medium is
  	\begin{equation}
		\rho \frac{\partial^2 u_i}{\partial t^2} = C_{ijkl} \frac{\partial^2 u_l}{\partial x_j \partial x_k}
	\label{eqn:fullWaveEquation},
	\end{equation}
    where the indices $i,j,k,l \in \{1,2,3\}$, $x$ and $t$ are the spatial and temporal variables and Einstein summation notation is used. The variable $u_i$ is the displacement in direction $i$. The elastic properties of each material are described by the density $\rho$ and the stiffness-tensor $C_{ijkl}$. The stiffness tensor has $81$ components which can be reduced to $21$ independent coefficients to describe a fully-anisotropic medium \cite{nayfeh1995wave}. Here, the analysis is restricted to materials that are either isotropic or transverse-isotropic, which reduces the number of independent coefficients further. As described previously, the wave-vectors of the partial-waves lie in a plane of material symmetry, $(x_1, x_3)$. Here $\partial/\partial x_2 = 0$ and the expanded form of the wave-equation Eq.\,\eqref{eqn:fullWaveEquation} is
    \begin{align}
    	\rho \frac{\partial^2 u_1}{\partial t^2}    &=  
    	C_{11} \frac{\partial^2 u_{1}}{\partial x_1^2} + C_{55} \frac{\partial^2 u_{1}}{\partial x_3^2} + (C_{13} + C_{55})\left( \frac{\partial^2 u_3}{\partial x_1 \partial x_3}\right) \nonumber
    	\\
    	\rho \frac{\partial^2 u_2}{\partial t^2}    &= 
    	C_{66} \frac{\partial^2 u_{2}}{\partial x_1^2} + C_{44} \frac{\partial^2 u_{2}}{\partial x_3^2} \nonumber
    	\\
    	\rho \frac{\partial^2 u_3}{\partial t^2}    &= 
    	C_{55} \frac{\partial^2 u_{3}}{\partial x_1^2} + C_{33} \frac{\partial^2 u_{3}}{\partial x_3^2} + (C_{13} + C_{55})\left( \frac{\partial^2 u_1}{\partial x_1 \partial x_2}\right),
    	\label{eqn:expandedWaveEquation}
    \end{align}
    where Voigt notation has been used to contract the indices of the stiffness-matrix (where $11 \rightarrow 1, 22 \rightarrow 2, 33 \rightarrow 3, 23 \rightarrow 4, 13 \rightarrow 5, 12 \rightarrow 6$). A single-frequency plane wave can be written in the form
    \begin{equation}
        \label{eqn:generalSolution}
         u_i = A_i \exp(\mathsf{i}(\zeta x_1 + \zeta \alpha x_3 - \omega t)),
    \end{equation}
    where $i \in \{ 1,2,3 \}$, $\omega$ is the circular frequency, $\alpha$ is the ratio of the vertical and horizontal ($\zeta$) wavenumbers, and $A_i$ is the polarisation unit vector which describes the direction of displacement relative to the direction of wave propagation. Substituting Eq.\,\eqref{eqn:generalSolution} into Eq.\,\eqref{eqn:expandedWaveEquation} gives the Christoffel equation
    \begin{equation}
        \Gamma_{ij}(\alpha) A_j = 0,
        	\label{eqn:ChristoffelForm}
    \end{equation}
    where the components of the Christoffel matrix ($\Gamma$) are
    \begin{align*}
    \Gamma_{11} &= (C_{11}  - \rho \nu^2 + C_{55} \alpha^2)  & \Gamma_{22} &= (C_{66} - \rho \nu^2 + C_{44} \alpha^2)  \\
    \Gamma_{33} &= (C_{55} - \rho \nu^2 + C_{33} \alpha^2) & \Gamma_{13} &= \Gamma_{31} =  (C_{13} + C_{55}) \alpha \\
     \Gamma_{12} &= \Gamma_{21} = \Gamma_{32} =\Gamma_{23} = 0. 
    \end{align*}
    The phase velocity $\nu$ along the $x_1$ axis is calculated from the relation $\nu = \omega/\zeta$. Solving Eq.\,\eqref{eqn:ChristoffelForm} admits three solutions for $\alpha^2$ and therefore $6$ solutions for $\alpha$.  From here, the notation $\alpha_q$, where $q \in \{1,2,...,6 \}$, will be used to indicate each solution.
   
   It can be seen from Eq.\,\eqref{eqn:ChristoffelForm} that the plane wave component $A_2$ is only dependent on $\Gamma_{22}$, hence displacement occurring in the $(x_1, x_3)$ plane is independent of displacement in $x_2$. 
   Four solutions ($q = 1,2,3,4$) of $\alpha_q$ Eq.\,\eqref{eqn:ChristoffelForm} can be found when
    \begin{equation}
    \label{eqn:christoffel2D}
    \det \left|
        \begin{array}{cc}
           \Gamma_{11}(\alpha)  & \Gamma_{13}(\alpha) \\
            \Gamma_{13}(\alpha) & \Gamma_{33}(\alpha)
        \end{array} \right| = 0.
    \end{equation}
    These describe upwards and downward travelling quasi-shear-vertical (qSV) and quasi-longitudinal (qL) waves with the displacement restricted to the $(x_1,x_3)$ plane.
    The remaining two solutions ($q = 5,6$) are found from $\Gamma_{22}=0$, and correspond to upward- and downward- travelling quasi-shear-horizontal (qSV) waves. 
    The notation $A_{iq}$ will be used to indicate polarisation vector for each solution $q$. 
    The displacement field can now be written as
    \begin{equation}
        u_i = \sum_q A_{iq} B_{q} \exp(\mathsf{i}(\zeta x_1 + \zeta \alpha_q x_3 - \omega t)),
    \end{equation}
    where $B_q$ is the amplitude of each partial-wave. Additionally, the stress field within the unbounded medium can be found using Hooke's law
    \begin{equation}
        \sigma_{ij} = C_{ijkl} \left( \frac{\partial u_k}{\partial x_l} + \frac{\partial u_l}{\partial x_k}  \right),
    \end{equation}
   	where $i,j,k,l \in \{1,2,3\}$. For a multi-layered medium, the Christoffel equation Eq.\,\eqref{eqn:ChristoffelForm} is solved independently for every layer to calculate the polarisation vector and wave-vector of each partial-wave. However, the amplitude $B_q$ of each partial-wave is solved by invoking the boundary conditions at the interfaces of adjacent layers. This is discussed in the following section.
   	
   \subsection{Boundary Conditions and Partial-wave Amplitudes}
   \label{section:layered}
    As mentioned previously, the wave-vector of the plane waves are constrained to a plane of symmetry of the transverse-isotropic material reducing the analysis to two dimensions, ($x_1, x_3$). The normal and transverse displacement and stress describing (quasi-)longitudinal and (quasi-)shear-vertical waves for a single layer is written in the form 
   \begin{equation} 
	    \left[ \begin{array}{c} u_1 \\ u_3 \\ \sigma_{33} \\ \sigma_{13} \end{array} \right] =
	    \left[ \begin{array}{cccc}
	    A_{11} & A_{12} & A_{13} & A_{14}  \\
		A_{31} & A_{32} & A_{33} & A_{34} \\
		D_{11} & D_{12} & D_{13} & D_{14} \\
		D_{21} & D_{22} & D_{23} & D_{24} 
	    \end{array} \right]
	     \left[ \begin{array}{cccc}
	    e_1 &  &  &  \\
	     & e_2 &  &  \\
	     &  & e_3 &  \\
	     &  &  & e_4 
	    \end{array} \right]
	    \left[  \begin{array}{c} B_1 \\ B_2 \\ B_3 \\ B_4 \end{array} \right],
	    \label{eqn:matrixForm}
    \end{equation}
     where 
    \begin{align*}
    D_{1q} &= (C_{13}A_{1q} + C_{33} \alpha_q A_{3q} )(\mathsf{i}\zeta) & D_{2q} &= C_{55}(\alpha_qA_{1
    q}+  A_{3q})(\mathsf{i} \zeta), \\ \nonumber
    e_q &= \exp(\mathsf{i} (\zeta x_1 + \zeta \alpha_q x_3 - \omega t)). 
    \end{align*}
    Only the first four solutions of $\alpha_q$ are needed as the motion is restricted to two dimensions ($x_1, x_3$).
    The left hand vector of Eqs.\,\eqref{eqn:matrixForm} contains the components of the displacement and stress, and the right hand vector contains the amplitude of the partial-wave components. The product of the matrices in Eq.\,\eqref{eqn:matrixForm} will be written as a field matrix $\mathbf{F}$. At an interface $x_3 = d$ between material layers in welded contact, the normal transverse stress and displacement must be continuous across the interface. 
    Therefore, the product of the field matrix and wave amplitudes at the interface of one layer is set equal to the field matrix and wave amplitudes of the adjacent layer. This process is repeated for every interface of the layered medium. For $n$-layers, there are $4(n-1)$ boundary conditions and $4n$ wave amplitudes which can be arranged into a global matrix. For example, for a medium consisting of $4$ layers, the global matrix equation is written
    \begin{equation}
    \label{eqn:globalMatrix}
         \left[ \begin{array}{cccc}
	       \mathbf{ F_1^1} & \mathbf{-F_2^1}  &  &  \\
	     & \mathbf{F_2^2} & \mathbf{- F_3^2}  &  \\
	     &  & \mathbf{F_3^3} & \mathbf{-F_4^3}  \\
	    \end{array} \right]
	    \left[  \begin{array}{c} \mathbf{B_{1}} \\ \mathbf{B_{2}} \\ \mathbf{B_{3}} \\  \mathbf{B_{4}}  \end{array} \right] = 0.
    \end{equation}
  	Here, $\mathbf{F}_n^N$ is the $4 \times 4$ field matrix and $\mathbf{B}_n$ a $4\times1$ vector of partial-wave amplitudes of layer $n$ at interface $N$. By assigning values to four of the partial-wave amplitudes, Eq.\,\eqref{eqn:globalMatrix}, can be rearranged and solved for the remaining partial-wave amplitudes.
  	For example, if a compressional wave in the first medium is incident on the layered-structure, the downward-travelling partial-wave amplitude relating to shear ($B_2^1$, Figure \ref{fig:workingDiagram}) in the first layer and upward-travelling partial-wave amplitudes relating to compressional ($B_3^n$) and shear ($B_4^n$) waves in the last ($n$th) layer are set to zero. Finally, the downward-travelling partial-wave amplitude relating to a compressional wave in the first layer is set to an arbitrary value ($B^1_1 = 1$). In this case, the solved amplitudes describe the solution for an incident single-frequency plane-wave at an angle $\theta$ or wavenumber $\zeta$ and frequency $f$. For a four-layered medium this would be 
  	    \begin{equation}
    \label{eqn:globalMatrixAxB}
         \left[ \begin{array}{cccc}
	       \mathbf{F_1^{1+}}& \mathbf{-F_2^1}  &  &  \\
	     & \mathbf{F_2^2} & \mathbf{- F_3^2}  &  \\
	     &  & \mathbf{F_3^3} & \mathbf{-F_4^{3-}}  \\
	    \end{array} \right]
	    \left[  \begin{array}{c} \mathbf{B_{1}^{+}} \\ \mathbf{B_{2}} \\ \mathbf{B_{3}} \\  \mathbf{B_{4}^{-}}  \end{array} \right] =      \left[ \begin{array}{cccc}
	       \mathbf{ -F_1^{1-}} &  &  &  \\
	     & &   &  \\  &  &  & \mathbf{F_4^{3+}}   \\
	    \end{array} \right]
	    \left[  \begin{array}{c} \mathbf{B_{1}^{-}} \\  \\  \\  \mathbf{B_{4}^{+}}  \end{array} \right]  .
    \end{equation}
    Here, $+$ and $-$ superscripts indicate the upwards and downwards travelling partial-wave amplitudes and their respective columns in the field matrices. For example, $\mathbf{F_1^{1+}}$ would be the third and fourth columns of $\mathbf{F_1^1}$, and $\mathbf{B^+_1}$ would be the third and fourth elements of $\mathbf{B_1}$. In the example described above, the first element of $\mathbf{B_1^-}$ is $1$  and all the elements of $\mathbf{B_4^+}$ and the second element of $\mathbf{B_1^-}$ are $0$.
  	Additionally, once the remaining wave-amplitudes for each layer are found, Eq.\,\eqref{eqn:matrixForm} can be used to find the displacement and stress anywhere in the layered structure. Alternatively, the dispersion curves can be extracted from the model by setting the incident wave-amplitudes of the layered structure to zero and finding the frequency-wavenumber pairs in which the resulting left-hand-side matrix of Eq.\,\eqref{eqn:globalMatrixAxB} becomes singular. The algorithm used is described further in Section \ref{section:implementationDetails} 
  	
  	\subsection{Shear-Horizontal Waves}
    Shear-horizontal waves propagate independently of (quasi-)shear-vertical and (quasi-)compressional waves and their propagation is analogous to compressional waves in a liquid \cite{brekhovskikh2012acoustics}. Solutions 5 and 6 of $\alpha_q$ correspond to shear-horizontal waves and the displacement and shear stress can be written in the form
    \begin{equation}
        \left[ \begin{array}{c}
             u_2\\ \sigma_{23} 
        \end{array} \right] = 
        \left[ \begin{array}{cc}
            A_{25} & A_{26} \\
            D_{35} & D_{36} 
        \end{array} \right]
        \left[ \begin{array}{cc}
           e_5  &   \\
             & e_6
        \end{array} \right]
        \left[ \begin{array}{c}
         B_5 \\ B_6
        \end{array} \right],
        \label{eqn:shmatrix}
    \end{equation}
    where
    \begin{equation}
        D_{3q} = C_{44} \alpha_{q} A_{2q} (i\zeta). \nonumber
    \end{equation}
    By setting the properties of a new medium $C'$ so that $C'_{11}, C'_{22}, C'_{33} = C_{44} $ and setting the remaining coefficients to zero, Eq.\,\eqref{eqn:matrixForm} reduces to Eq.\,\eqref{eqn:shmatrix}. Hence, the propogation of shear-horizontal waves can be modelled with Eq.\,\eqref{eqn:matrixForm} by reassigning the relevant components of the stiffness matrix.

 	\subsection{Implementation Details}
 	\label{section:implementationDetails}
 	To construct the global matrix, Eq.\,\eqref{eqn:globalMatrix}, a field matrix $\mathbf{F_i}$ must be calculated for each layer. To improve the conditioning of the matrix, rows relating to displacement equations are scaled by $\zeta$ the horizontal wavenumber, and rows relating to stress are scaled by $\rho \omega^2$. The system matrix is constructed by looping over each interface, calculating the $4\times4$ field matrices above and below each interface and arranging them into a single matrix. This leads to a rectangular matrix which has $4n$ columns and $4n-1$ rows. Additionally there are $4n$ partial-wave amplitudes. Four partial-wave amplitudes are defined and the global-matrix in Eq.\,\eqref{eqn:globalMatrix} is rearranged to be square Eq.\,\eqref{eqn:globalMatrixAxB}. The resulting equation is solved using the \texttt{mldivide} function in MATLAB. This function solves a system of linear equations using the fastest algorithm based on the matrix structure. However, the global-matrix becomes singular at values of $\zeta$ and $\omega$ on or close-to dispersion curve solutions.
  	
  	The computation of dispersion curves follows the algorithm described in \cite{Lowe1995a}. Firstly, the wavenumber parameter is fixed and the determinant of the global-matrix is found over a range of frequencies. Close to dispersion solutions the determinant of the global-matrix tends to zero. Using these as starting points and taking a limit either side, the exact frequency and wavenumber of the dispersive solution is found using a bisection algorithm. \texttt{ElasticMatrix} makes use of MATLAB's \texttt{fmincon()} function for this. These solutions are the starting points for each dispersion curve. To find the second point on each dispersion curve, the fixed value of wavenumber is increased and the search is performed again. The algorithm then uses linear interpolation to estimate the location of the third, fourth and fifth points on the dispersion curve, similarly using a bisection algorithm to find the exact frequency-wavenumber pairs. After five points have been found, a higher-order polynomial interpolation scheme is used to more accurately predict points on the dispersion curve. The algorithm implemented in \texttt{ElasticMatrix} only searches in the real domain of $\zeta$ which is a good-estimate for simple plate structures in a vacuum, however, it may be inaccurate for leaky solutions, for example an plate embedded in soil.

    Slowness profiles are calculated by defining a range of phase-speeds and calculating the horizontal and vertical component of wavenumber by solving the Christoffel equation Eq.\,\eqref{eqn:christoffel2D}. The values from the calculation may be complex, however, only the real values are plotted.


\section{Software Description and Examples}

\label{section:softwareDescription}
    \subsection{Overview}
\begin{figure}
        \centering
        \includegraphics[width=0.96\textwidth]{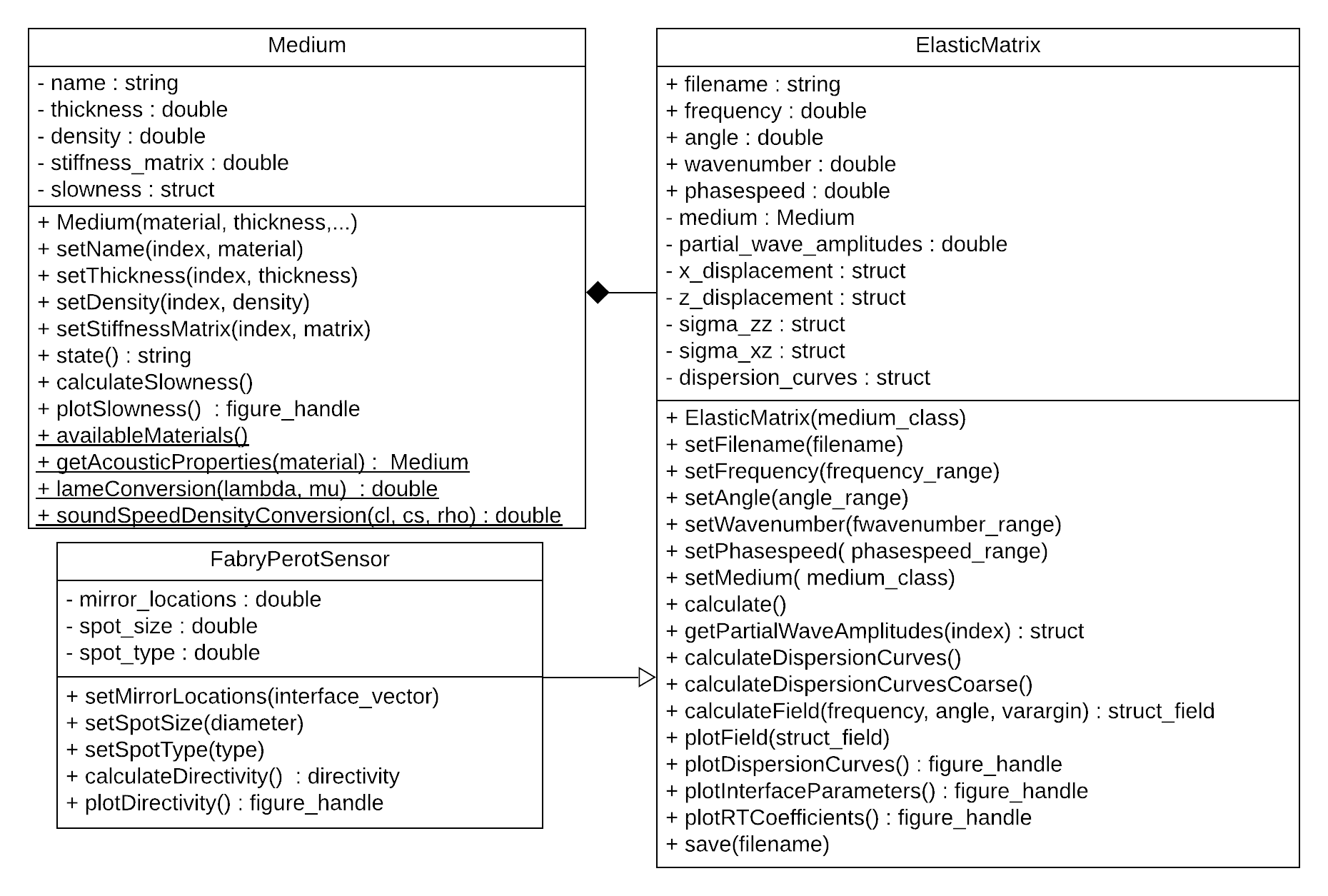}
        \caption{UML class diagram for \texttt{Medium}, \texttt{ElasticMatrix} and \texttt{FabryPerotSensor}. The top field for each box indicates the name of the class, the second field lists the properties, and the third field lists the methods. Here, \texttt{ElasticMatrix} is composed from \texttt{Medium} and \texttt{FabryPerotSensor} inherits \texttt{ElasticMatrix}. The $(-)$ indicates a private method or property and $(+)$ indicates a public method or property. Underlined methods are static.}
        \label{fig:classDiagram}
    \end{figure}

    The \texttt{ElasticMatrix} toolbox implements the partial-wave method using an object-oriented framework in MATLAB. This allows the toolbox to be used with either a simple scripting or command line interface, and makes it easy to use and expand. The software is divided into three classes. The first class, \texttt{Medium}, defines the multi-layered geometry and material properties of each layer. The second class \texttt{ElasticMatrix} is initialised by a \texttt{Medium} object. This class contains the partial-wave method implementation and methods for extracting additional details such as dispersion curves and reflection coefficients. By default, all the calculations use 64 bit precision. The final class, \texttt{FabryPerotSensor}, is an example of how  numerical models can be built from the \texttt{ElasticMatrix} and \texttt{Medium} objects. This class inherits \texttt{ElasticMatrix} and can be used to model the directional response of a Fabry-P\'erot ultrasound sensor. More details can be found in \cite{ramasawmy2018directivity, cox2007frequency}. Each class in the toolbox inherits the MATLAB handle class. Consequently, the object does not need to be reassigned when a method is called. The classes and their respective attributes and methods can be seen in Fig. \ref{fig:classDiagram}. The toolbox is self contained and has been tested with MATLAB 2016a and above.
    
    This section presents a small selection of code snippets and examples. More detailed examples can be found in the \texttt{ElasticMatrix} \texttt{./examples} folder and \texttt{html} documentation can be accessed through the MATLAB help and clicking \texttt{ElasticMatrix toolbox}. There are three steps to using the toolbox. Firstly, the geometry of the layered medium must be defined. Secondly, the input parameters to the model should be defined, which are generally a range of angles, frequencies or wavenumbers. Finally, the model can be solved and details such as the reflection coefficients and dispersion curves can be extracted. Note, for clarity in the code implementation, the $x_1$ and $x_3$ coordinates are referred to as $x$ and $z$, respectively.
    
    \subsection{\texttt{Medium}}
    The \texttt{Medium} class is used to define the material properties and thickness of each layer. The class is initialised by calling the class constructor with input arguments of the material name followed by its thickness. However, the thickness of the first and last layers are semi-infinite and their values should be set with the \texttt{Inf} keyword. The \texttt{Medium} class will automatically set the thickness of the first and last layer to \texttt{Inf} if another value is used. An example is given below.
        \begin{verbatim}
my_medium = Medium(`water', Inf, `blank', 3e-3, `PVDF', 1e-3, `glass', Inf);  \end{verbatim}
    Here, \texttt{my\_medium} is an object array and every index in the object array corresponds to a different layer in the medium. In the current example, \texttt{my\_medium(3)} will return a object with the material properties and thickness associated with PVDF. The \texttt{`blank'} keyword can used for a material which is not predefined. The material properties and names can be set using their respective \texttt{set} functions. User defined materials can be added to the script \texttt{materialList.m}. 
	\subsection{Slowness profiles}
\begin{figure}[ht]
        \centering
        \includegraphics[width=\textwidth]{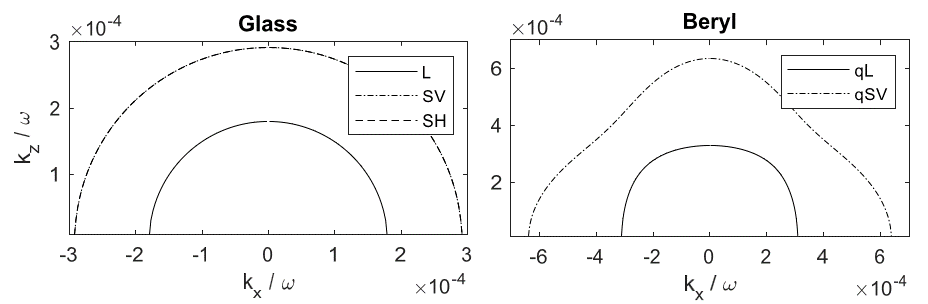}
        \caption{The slowness curves for isotropic-glass and transverse-isotropic beryl materials where (q)L, (q)SV, (q)SH correspond to the reciprocal of the (quasi-)longitudinal, (quasi-)shear-vertical and (quasi-)shear-horizontal partial-wave speeds.}
        \label{fig:slownessProfile}
    \end{figure}

	Slowness profiles are a plot of the inverse-phase velocity of each bulk wave component. They can be used to determine the angles of reflection and transmission between multi-layered media as well as the direction of energy propagation and skew angle \cite{Rose2004}. Slowness profiles are found by solving the Christoffel equation, Eq. \ref{eqn:ChristoffelForm}, and only depend on the material properties of each material. The method \texttt{.calculateSlowness} is part of the \texttt{Medium} class, and calls the function 	%
  	\begin{verbatim} calculateAlphaCoefficients(...) \end{verbatim} 
  	which is an implementation of Eqs.\,\eqref{eqn:ChristoffelForm} and \eqref{eqn:christoffel2D}. This takes input arguments of the material properties and phase-velocity and returns the polarisation and wave-vectors. The slowness profiles given by this function are plotted in terms of $k_x/\omega$ vs $k_z/\omega$. For an isotropic material, the slowness profiles for each bulk wave are spherical, however, this is not true for an anisotropic material. An example of the slowness profiles for isotropic-glass and transverse-isotropic beryl is shown in Fig. \ref{fig:slownessProfile}. This figure has been reproduced from \cite{payton2012elastic}. The slowness profiles of the (quasi-)longitudinal, (quasi-)shear-vertical and (quasi-)shear-horizontal bulk waves are shown. As glass is an isotropic material, the slowness profiles are spherical and the magnitudes of L,SV and SH when $k_x/\omega = 0$ or $k_z/\omega = 0$ are equal to the reciprocal of the compressional- and shear-speeds of glass. For the transverse-isotropic case, when $k_x/\omega = 0$, the value of $qL$ is equal to $\sqrt{\rho/C_{33}}$ and $qSV$ is equal to $\sqrt{\rho/C_{55}}$. When $k_z/\omega = 0$, the value of $qL$ is equal to $\sqrt{\rho/C_{11}}$ and the value of $qSV$ is equal to $\sqrt{\rho/C_{55}}$. These have been checked in the toolbox example script and all are within numerical precision.
	\begin{verbatim}
my_medium = Medium(`glass', Inf, `beryl', Inf);
my_medium.calculateSlowness;
my_medium.plotSlowness;
	\end{verbatim}
  	\subsection{\texttt{ElasticMatrix}}
    The medium class is used to initialise the \texttt{ElasticMatrix} class which runs the partial-wave method over a range of frequencies, wavenumbers, phasespeeds and angles. Two of these must be defined using the \texttt{.set} functions. The \texttt{.calculate} method is then used to run the partial-wave procedure. If the properties are not set before the \texttt{.calculate} method is called, the model will run over a predefined range of frequencies and angles. 
    The \texttt{.calculate} method constructs, rearranges and solves the global-matrix, Eq.\,\eqref{eqn:globalMatrixAxB}, using the function
  	\begin{verbatim} calculateMatrixMethod(...) 	\end{verbatim}
  	This function takes input arguments of the material properties and the parameters to calculate over (angles, frequencies, wavenumbers). It returns the determinant of the system matrix and the stresses and displacements at the layer interfaces. Each individual field-matrix is calculated using the function  
  	\begin{verbatim} calculateFieldMatrixAnisotropic(...)	\end{verbatim}
  	which is an implementation of Eq.\,\eqref{eqn:matrixForm}. This takes input arguments of the material properties, the wave-vector components, polarisation components and the phase velocity and returns the field-matrix. The default calculation is to find the partial-wave amplitudes and interface stresses and displacements when there is a single-frequency compressional wave incident on the structure from the first layer. An example is given below for a titanium plate.
    \begin{verbatim}
my_medium = Medium(`water', Inf, `titanium', 1e-3, `water', Inf);
my_model = ElasticMatrix(my_medium); % initialise class
my_model.setFrequency(linspace(1e6, 5e6, 100));
my_model.setAngle(linspace(0, 45, 100));
my_model.calculate; \end{verbatim}
	\subsection{Reflection and Transmission Coefficients}
	For a plane wave incident at an oblique angle on a multi-layered structure, the reflection and transmission coefficients determine the amplitude of the wave that is reflected and transmitted at each interface. Knowing these coefficients is useful for a number of applications. For example, selecting the appropriate launch angle when coupling energy into particular modes in a wave-guide, or determining the thickness and material properties of matching layers for ultrasonic transducers \cite{Rose2004, kelly2004characterization}. 
	
	The angles of refraction at the interfaces between multi-layered media can be found by studying the slowness profiles. However, slowness profiles do not take into account the boundary conditions at the interfaces. Consequently, the magnitude of each of the refracted waves cannot be calculated directly. For a plane wave incident on a multi-layered structure, the magnitude of the reflection and transmission coefficients are found by normalising the partial-wave amplitudes $B_i^n$ by the incident plane wave amplitude $B_1^1$. This is automatically calculated when using the \texttt{.calculate} method. 
	
\begin{figure}[ht]
        \centering
        \includegraphics[width=0.6\textwidth]{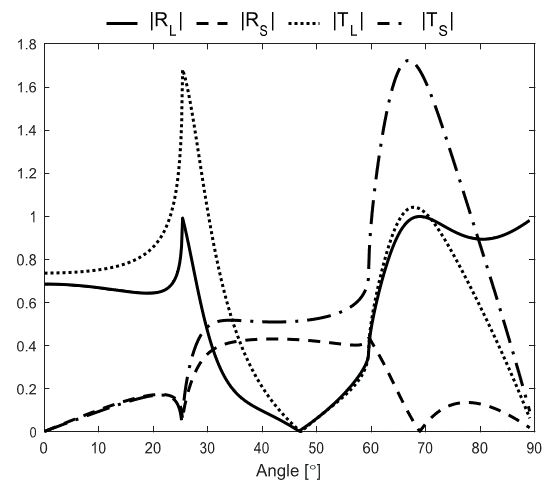}
        \caption{Longitudinal $L$ and shear $S$ reflection $R$ and transmission $T$ coefficients for a PVDF-Aluminium interface. }
        \label{fig:reflectionCoefficient}
    \end{figure}

	An example of the reflection and transmission coefficients at a PVDF-aluminium interface is given below and shown in Fig. \ref{fig:reflectionCoefficient}. For a plane compressional wave incident on a PVDF-aluminium interface, there are four resulting refracted waves. These are a reflected $R$ and transmitted $T$ compressional $L$ and shear $S$ wave. The reflection and transmission coefficients have been compared to the analytic solutions for a two-layered elastic-medium from \cite{Rose2004} and have an average error less than $1e^{-15}$ which is within numerical precision for a 64 bit floating point number. For clarity the analytical solutions have not been plotted but can be seen in the toolbox examples folder.
	\begin{verbatim}
my_medium = Medium(`PVDF', Inf, `aluminium', Inf);
my_model = ElasticMatrix(my_medium);
my_model.setFrequency(1e6);
my_model.setAngle(linspace(0, 90, 90));
my_model.calculate;
my_model.plotRTCoefficients;	\end{verbatim}
	\subsection{Dispersion Curves}
	Dispersion curves describe the modal solutions of the multilayer structure and describe a wave-mode which propagates parallel to the layer-interfaces independently of a bulk wave. As one example, exciting these modes is essential in ultrasonic inspection. Knowledge of the dispersion curves is useful for determining the most appropriate modes to excite and for optimising the inspection process. 
	
\begin{figure}[ht]
        \centering
        \includegraphics[width=\textwidth]{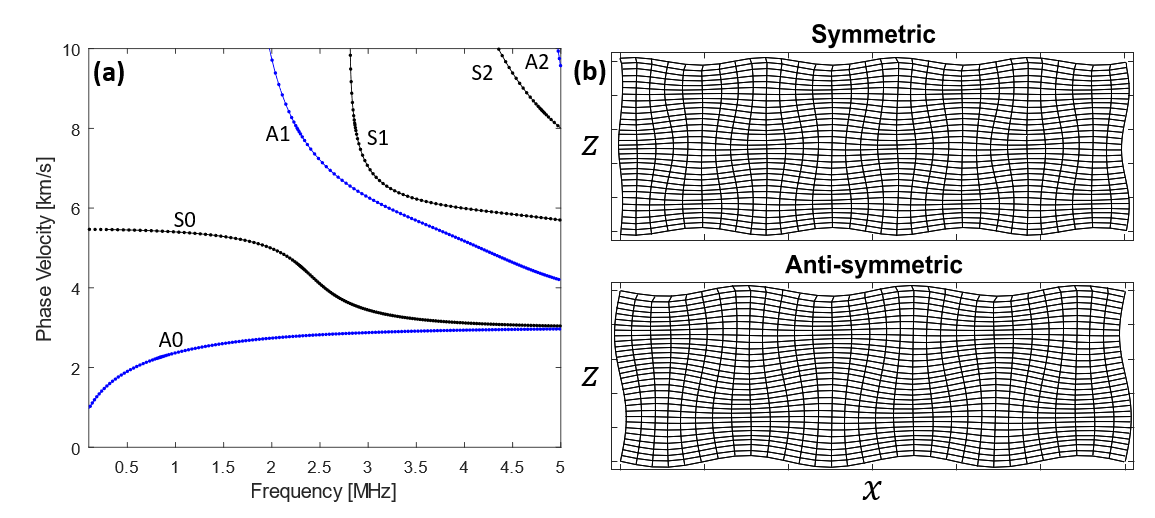}
        \caption{(a) Dispersion curves for a titanium plate in a vacuum. The solid lines are from \texttt{ElasticMatrix} and the points are generated using Disperse \cite{pavlakovic1997disperse}. The first three symmetric (S, black) and anti-symmetric (A, blue) have been plotted. (b) Displacement field for an anti-symmetric and symmetric mode shape.}
        \label{fig:dispersionCurves}
    \end{figure}

	As mentioned in Section \ref{section:implementationDetails}, the modal solutions are found when the global matrix becomes singular. The \texttt{ElasticMatrix} software can calculate dispersion curves for simple layered structures (i.e., a plate in a vacuum or water). However, it is not robust for very-leaky cases, for example a plate embedded in soil. For these types of cases either Disperse, or other techniques based on the spectral-collocation method or semi-analytic finite element method are more appropriate \cite{pavlakovic1997disperse,quintanilla2016full, quintanilla2015guided, quintanilla2015modeling, adamou2004spectral,sorohan2011extraction}. An example of the dispersion curves for a $1$\,mm titanium plate in a vacuum is shown in Fig. \ref{fig:dispersionCurves}(a). The dispersion curves are plotted on a graph of frequency vs wavenumber and show the first three symmetric $S$ and anti-symmetric $A$ Lamb modes. The results from Disperse are also plotted and have excellent agreement.
		\begin{verbatim}
my_medium = Medium(`vacuum', Inf, `titanium', 0.001, `vacuum', Inf);
my_model = ElasticMatrix(my_medium);
my_model.setFrequency(linspace(0.5e6, 5e6, 100));
my_model.calculateDispersionCurves;
my_model.plotDispersionCurves;
	\end{verbatim}
	 
\subsection{Displacement and Stress Fields}
    More information about the wave-physics and guided wave structures can be taken from dispersion curves by plotting the displacement and stress fields at different points. In the \texttt{ElasticMatrix} software implementation, the $x$ and $z$ ranges over which to plot the displacement or stress fields must be specified. The \texttt{.calculateField(...)} method returns a structure with the input ranges and field values at each point of the resulting grid. The values of the structure can be plotted independently or given as an argument to the \texttt{.plotField} method. An example is given below for the displacement field within an titanium plate for a symmetric and anti-symmetric mode. The resulting plot can be seen in Fig. \ref{fig:dispersionCurves}(b). 
\begin{verbatim}
field_values = myModel.calculateField(freq, angle, {x_range, z_range});
myModel.plotField(field_values, plot_style);
\end{verbatim}

    \subsection{\texttt{FabryPerotSensor}}
\begin{figure}[!ht]
    \centering
    \includegraphics[width=0.7\textwidth]{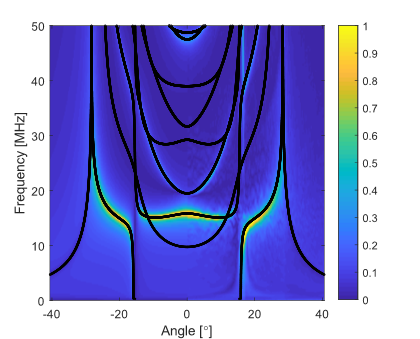}
    \caption{ The modelled directional response ($-40^{\circ}-0^{\circ}$) and measured directional response ($0^{\circ}-40^{\circ}$) from \cite{ramasawmy2018directivity}. The dispersion curves associated with this sensor are plotted as black points. }
    \label{fig:FabryPerot}
\end{figure}

	One application of the toolbox is to model the directional response of Fabry-P\'erot ultrasound sensors \cite{ramasawmy2018directivity}.  The \texttt{FabryPerotSensor} is a child class of \texttt{ElasticMatrix} and is an example of how the \texttt{ElasticMatrix} toolbox may be expanded. The are additional inputs to this class which are described in more detail in the \texttt{./examples} folder, and a description of the modelling process can be found in \cite{ramasawmy2018directivity, cox2007frequency}. The modelled directional response was found to have good agreement with the measured directional response. An example of the modelled and measured directional response for a glass etalon Fabry-P\'erot sensor can be seen in Fig. \ref{fig:FabryPerot}. Features of the directional response correspond to symmetric and anti-symmetric Lamb modes propagating within the sensor. 
	\begin{verbatim}
my_medium = Medium(`water', Inf, `AlMir', 1e-8, `glass', 175e-6, `AlMir', ...
    1e-8, `air', Inf);
fp_sensor = FabryPerotSensor(my_medium);
fp_sensor.setAngle(linspace(0, 45, 45));
fp_sensor.setFrequency(linspace(0.1e6, 100e6, 100));
fp_sensor.setMirrorLocations([1, 4]);
fp_sensor.calculateDirectivity;
fp_sensor.plotDirectivity;
fp_sensor.calculateDispersionCurves;
\end{verbatim}
    \squeezeupt

\section{Impact and Conclusions}
\label{section:impact}
This paper introduces a new open-source toolbox called \texttt{ElasticMatrix} which models elastic wave propagation in multi-layered media with anisotropic materials with isotropic or transverse-isotropic symmetry. The toolbox uses the partial-wave method which allows the calculation of slowness profiles, reflection and transmission coefficients, dispersion curves and stress and displacement fields. The software has been implemented using the object-oriented capabilities of MATLAB allowing for a simple command line or scripting interface. The implementation allows researchers to add functionality and integrate the software into other projects. For example, this toolbox has already been used to model the directional response of Fabry-P\'erot ultrasound sensors \cite{ramasawmy2018directivity}. It is anticipated the research user-base will actively contribute to \texttt{ElasticMatrix} and add to the functionality. 

\section*{Acknowledgements}
This work was supported in part by the Engineering and Physical Sciences Research Council (EPSRC), UK, Grant Nos. EP/P008860/1, EP/L020262/1 and EP/S026371/1, in part by the EPSRC-funded UCL Centre for Doctoral Training in Medical Imaging (EP/L016478/1) and the Department of Health’s NIHR-funded Biomedical Research Centre at NIHR Biomedical Research Centre at University College London Hospitals.

\bibliographystyle{elsarticle-num} 
\bibliography{references}

\section*{Current code version}
\label{section:CodeVersion}

\begin{table}[!ht]
\begin{tabular}{|l|p{6.5cm}|p{6.5cm}|}
\hline
\textbf{Nr.} & \textbf{Code metadata description} & \\
\hline
C1 & Current code version & v1 \\
\hline
C2 & Permanent link to code/repository used for this code version & github.com/dannyramasawmy/ ElasticMatrix \\
\hline
C3 & Legal Code License   & GNU Lesser General Public License v3.0 \\
\hline
C4 & Code versioning system used & git \\
\hline
C5 & Software code languages, tools, and services used & MATLAB. \\
\hline
C6 & Compilation requirements, operating environments \& dependencies & MATLAB 2016a and above \\
\hline
C7 & If available Link to developer documentation/manual &  github.com/dannyramasawmy/ ElasticMatrix \\
\hline
C8 & Support email for questions & dannyramasawmy+elasticmatrix @gmail.com \\
\hline
\end{tabular}
\caption{Code metadata}
\label{table:Metadata} 
\end{table}

\end{document}